\documentclass[letterpaper]{sig-alternate-2013}
\usepackage{url}
\usepackage{graphics}
\usepackage{color}
\usepackage[usenames,dvipsnames]{xcolor} 
\usepackage{url}
\usepackage{subfigure}
\usepackage{booktabs}
\usepackage{tipa}

\usepackage[pdftex]{hyperref}
\hypersetup{
  pdftitle={A Computer-Based Method to Improve the Spelling of Children with Dyslexia},
  pdfauthor={Luz Rello, Clara Bayarri, Yolanda Otal, Martin Pielot},  
  pdfkeywords={Dyslexia; Literacy; Serious Game; Spelling; Written Errors},
  pdfsubject={Human Factors},
bookmarksnumbered,
pdfstartview={FitH},
colorlinks,
citecolor=black,
filecolor=black,
linkcolor=black,
urlcolor=black,
breaklinks=true,
}

\newfont{\mycrnotice}{ptmr8t at 7pt}
\newfont{\myconfname}{ptmri8t at 7pt}
\let\crnotice\mycrnotice%
\let\confname\myconfname%

\permission{Permission to make digital or hard copies of all or part of this work for personal or classroom use is granted without fee provided that copies are not made or distributed for profit or commercial advantage and that copies bear this notice and the full citation on the first page. Copyrights for components of this work owned by others than ACM must be honored. Abstracting with credit is permitted. To copy otherwise, or republish, to post on servers or to redistribute to lists, requires prior specific permission and/or a fee. Request permissions from Permissions@acm.org.}
\conferenceinfo{ASSETS'14,}{October 20--22, 2014, Rochester, NY, USA.}
\copyrightetc{Copyright 2014 ACM \the\acmcopyr}
\crdata{978-1-4503-2720-6/14/10\ ...\$15.00.\\
http://dx.doi.org/10.1145/2661334.2661373}

\begin{document}

\title{A Computer-Based Method to Improve the Spelling \\ of Children with Dyslexia}

\numberofauthors{3}

\author{
 \alignauthor Luz Rello \\ Clara Bayarri\\
       \affaddr{Cookie Cloud}\\
       \affaddr{Barcelona, Spain}\\
       \affaddr{luz,clara@cookie-cloud.com}
\alignauthor Yolanda Otal  \\
       \affaddr{Centre Creix}\\
       \affaddr{Barcelona, Spain}\\
       \affaddr{otal@creixbcn.com}  
\alignauthor Martin Pielot\\
    	\affaddr{Telefonica Research}\\
   	\affaddr{Barcelona, Spain}\\
    	\affaddr{martin.pielot@telefonica.com}
       }
\maketitle
	
\begin{abstract}
In this paper we present a method which aims to improve the spelling of children with dyslexia through playful and targeted exercises. In contrast to previous approaches, our method does not use correct words or positive examples to follow, but presents  the child a misspelled word as an exercise to solve. We created these training exercises on the basis of the linguistic knowledge extracted from the errors found in texts written by children with dyslexia. 
To test the effectiveness of this method in Spanish, we integrated the exercises in a game for iPad, {\it DysEggxia} ({\it Piruletras} in Spanish), and carried out a within-subject experiment. During eight weeks, 48 children played either {\it DysEggxia}  or {\it Word Search}, which is another word game. We conducted tests and questionnaires at the beginning of the study, after four weeks when the games were switched, and at the end of the study. The children who played {\it DysEggxia} for four weeks in a row had significantly less writing errors in the tests that after playing {\it Word Search} for the same time. This provides evidence that  error-based exercises presented in a tablet  help children with dyslexia improve their spelling skills.
\end{abstract}

\keywords{Dyslexia; Literacy; Serious Game; Spelling; Written Errors}

\category{K.4.2}{Computers and Society}{Social Issues}[Assistive technologies for persons with disabilities]
\category{K.3}{Computers in Education}{Computer Uses in Education}[Computer-assisted instruction].


\section{Introduction}
Worldwide, around 15-20\% of the population has a language based learning disability \cite{IDA}. Likely, 70-80\% of them have dyslexia \cite{IDA}, a neurological learning disability which impairs a person's ability to read and write.  Overcoming dyslexia means a great effort for children and requires doing regular language exercises \cite{Hornsby2011}. Traditionally, these exercises are done using pen and paper. More recently, it was shown that computer games are a convenient medium to provide exercises in an engaging way to significantly improve the reading performance of children with dyslexia \cite{Kyle2013,Lyytinen2007}. Regarding writing, there are some technologies to support writing such as spellcheckers and word prediction software for people with dyslexia. However, it has not been shown that these tools improve spelling skills. 

In this paper we present the first computer-based approach to improve the spelling skills for people with dyslexia. Since, the writing errors of people with dyslexia are related to the types of difficulties that they have \cite{Sterling1998}, we use real errors found in texts written by children with dyslexia to create training exercises. The exercises were integrated in a game, called {\it DysEggxia} ({\it Piruletras} in Spanish), that shows the player an incorrect word that has to be corrected. Since dyslexic readers cannot consciously see errors in words \cite{Bruck1988,WWW2012}, our hypothesis is that children could learn how to identify typical dyslexic errors and, therefore, develop compensating strategies to write better.



To evaluate our method we conducted an eight-week experiment with 48 children with dyslexia. We compared the evolution of reading and spelling skills using {\it DysEggxia} and a baseline condition, {\it  Word Search},  another word-exercise game for iPad.
After playing {\it DysEggxia} for four weeks, children made significantly less writing errors compared to the ones that played the control condition game. Our results provide evidence that exercises on the basis of errors allow children with dyslexia to improve their spelling skills.

Next, we describe related work, the design of our method and our hypotheses followed by the evaluation. Later, we present and discuss the results and we draw the conclusions. 

\section{Dyslexia}
Dyslexia is a specific learning disability with neurological origin. It is characterized by difficulties with accurate and/or fluent word recognition and by poor spelling and decoding abilities. Secondary consequences may include problems in reading comprehension and reduced reading experience that can impede growth of vocabulary and background knowledge \cite{Lyon2003}. Dyslexia is frequent. From 10 to 17.5\% of the population in the U.S.A. \cite{CommitteeonLearningDisabilities1987} and from 8.6\% \cite{Jimenez2009} to 11.8\% \cite{Carrillo2011} of the Spanish speaking population have this disability.

Since literacy acquisition is essential for all aspects of learning, high rates of academic failure are associated with dyslexia when it is not diagnosed and treated correctly \cite{Gabrieli2009}. Actually, the most frequent way to detect a child with dyslexia is by her or his low performance at school \cite{Carrillo2011}. For instance, in Spain, approximately four out of six cases of school failure are related to a language based learning disability \cite{FEDIS}.\footnote{The percentage of school failure is calculated by the number or students who drop school before finishing secondary education (high school). While the average of school failure in the European Union is around 15\%, Spain has around 25-30\% of school failure, 31\% in 2010 \cite{Enguita2010}.}

\section{Related Work}
We divide the related work in assistive technologies to support dyslexia and technologies for treating dyslexia.

\subsection{Assistive Technologies for Dyslexia}
\subsubsection{Reading} 

Previous work has shown that specific text presentations can make text easier to read \cite{Gregor2000,Kurniawan2006}.
Santana {\it et al}. \cite{W4A2013Santana} developed {\it Firefixia}, a tool that allows readers with dyslexia to customize websites to improve readability.
They tested {\it Firefixia} with four users and found that readers with dyslexia appreciate customization.
Dickinson {\it et al.} \cite{Dickinson2002} asked 12 students with dyslexia to test different colors, sizes, spacings, column widths, and letter highlighting to improve the subjective readability of MS Word documents integrated in {\it SeeWord} sorfware. The results were tested by seven people with dyslexia, which reported a subjective increase in readability.
{\it Text4All} \cite{Topac2012} for websites, the Android {\it IDEAL eBook reader} \cite{ASSETS2012}, and the iOS {\it DysWebxia Reader} \cite{ASSETS2013demo} are text customization tools developed on the basis of previous research using eye-tracking with people with dyslexia \cite{W4A2012}. Also, the multimodal {\it MultiReader} was designed on the basis of user studies for readers with dyslexia as well as blind, partially sighted, and deaf people~\cite{Petrie2001multireader}. 


\subsubsection{Writing}

Previous work has also explored how to help writing.
For example, tailoring spellcheckers to detect typical errors committed by people with dyslexia \cite{Korhonen2008,Li2013,Pedler2007}. These adaptations include the detection of real-word errors, such as {\it *witch} instead of {\it which} or boundary errors, such as {\it *miss spelled} instead of {\it misspelled}.\footnote{Examples with errors are preceded by an asterisk ``*''. We use the standard linguistic conventions: `\textless \textgreater' for graphemes,  `/ /' for phonemes and `[ ]' for allophones.} 
In addition, Khakhar and Madhvanath \cite{Khakhar2010} describe {\it JollyMate}, a tool that provides handwritten character recognition to detect miswritten characters.

We believe that these assistive technologies could make reading and writing easier, however, it has not been shown that they improve literacy skills.

\subsection{Technologies for Treating Dyslexia}

\subsubsection{Action Games} 
Franceschini {\it et al.}~\cite{Franceschini2013} investigated whether computer action games help children with dyslexia to improve their ability of decoding words. 
In a between-group experiment with 20 children, they observed that 10 children playing action games for 9 sessions (80 min each) improved their reading skills significantly more than the control group that played non-action games. They claim that action games improve the children's spatial and temporal attention, which is essential for decoding words.

\subsubsection{Serious Games to Identify Dyslexia}
Lyytinen {\it et al.}~\cite{Lyytinen2007} created the computer game {\it Literate}, later called {\it GraphoGame} \cite{Lyytinen2009},\footnote{\small\url{https://graphogame.com/}} which was developed to identify children at risk of having dyslexia before school age in Finland. Its exercises are aimed towards the connection of graphemes (letters) and phonemes (sounds) to improve reading.
They conducted two user studies with 12 and 41 children between 6 and 7 years old with very promising results. Children who used  {\it Literate} improved their accuracy in grapheme-phoneme connections, reading words, and naming phonemes after playing for less than 4 hours. These results show that linguistic-based exercises can support the improvement of reading.

%

\subsubsection{Computer-Assisted Reading Interventions}

\sloppy
Kyle {\it et al.} \cite{Kyle2013} compared two computer-assisted reading interventions for English inspired by the Finnish {\it GraphoGame}: {\it GG Rime} and {\it GG Phoneme}. In both games, the goal is to train grapheme-phoneme connections. The player hears either sounds or words and has to match them to visual targets (letters and sequences of letters) displayed on the screen. {\it GG Rime} includes rhyming word families to reinforce grapheme-phoneme connections.
They conducted a user study with 31 children of 6 and 7 years old. They were divided in three groups of 11, 10, and 10 students which were exposed to {\it GG Rime}, {\it GG Phoneme}, and no treatment, respectively, for 5 sessions (10-15 minutes each) per week during 12 weeks. 
While the results show that both games may benefit decoding abilities, no significant effects were found, probably due to an insufficient number of participants or not enough training time.

\subsection{What is Missing?}

A technology to train and improve spelling skills of people with dyslexia because: (1) previous studies for treating dyslexia were more focused on the development of reading skills such as grapheme-phoneme connections or the improvement of attention abilities in relationship with reading acquisition; and (2) to the best of our knowledge there are no scientific reports on any assistive or treating technology which improves writing acquisition in children with dyslexia.

\begin{figure}[tb!]
	\centering
	\includegraphics[width=8.5cm]{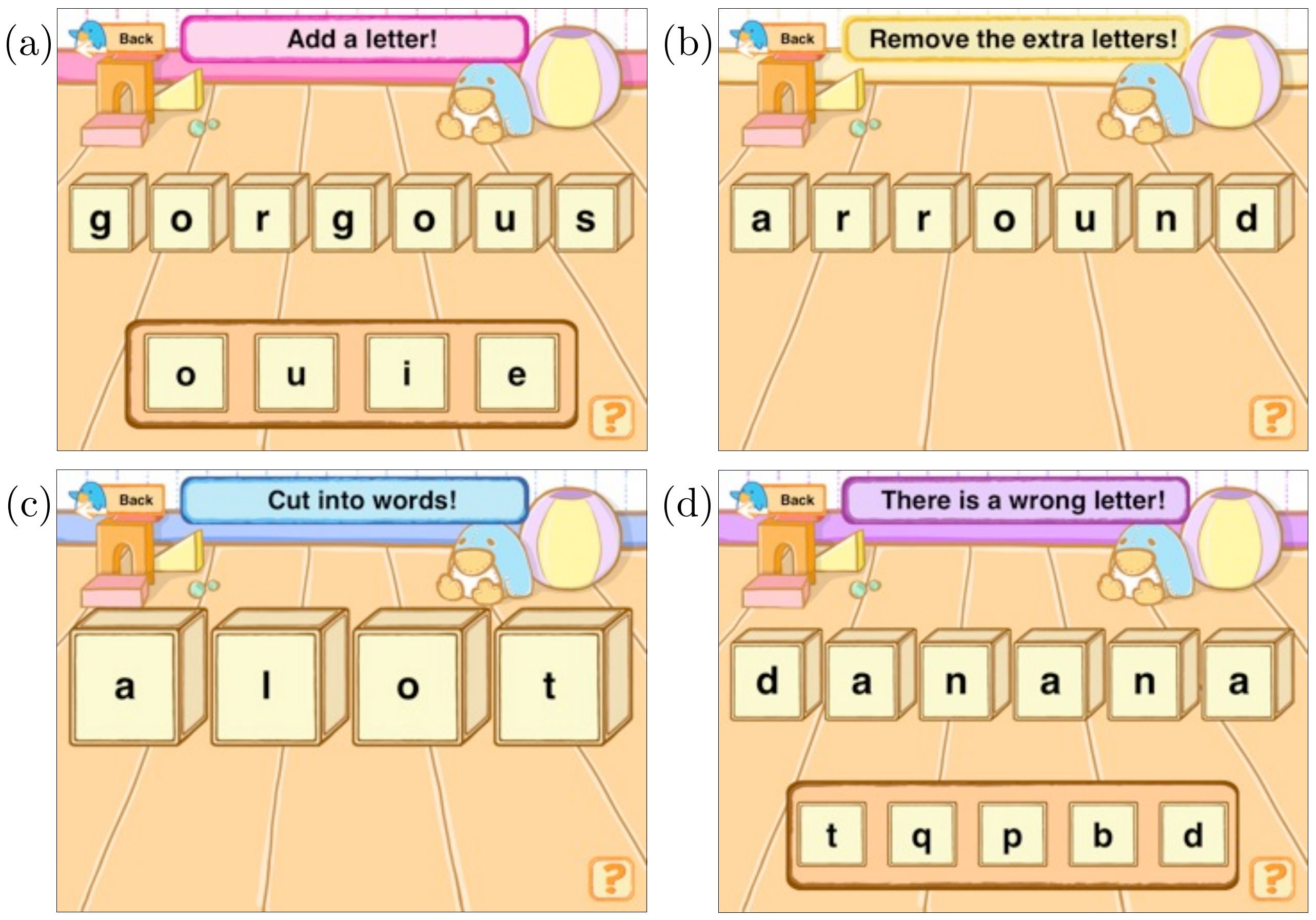}
		\vspace{-0.5cm}
	\caption{{\it DysEggxia} exercises of (a) {\it add a letter}, (b) {\it remove a letter}, (c) {\it cut into words}, and (d) {\it change a letter}.}
	\label{dysapp}
		\vspace{-0.3cm}
\end{figure}

\section{Game Design}
\label{app}
{\it DysEggxia} is a game to support the spelling acquisition of children with dyslexia through the realization of exercises.\footnote{{\it DysEggxia} was demoed at ASSETS'12 \cite{Dyseggxia2012}.} The goal of the exercises is to produce correct words. These were designed on the basis of the linguistic analysis of errors written by children with dyslexia.

We chose to use a touch interface as input media for the game instead of handwriting, because it allows to isolate the writing problems that are caused by dyslexia and not by dysgraphia \cite{Romani1999}.\footnote{Dysgraphia is a writing disorder associated with the motor skills involved in handwriting, sequencing, and orthographic coding. It is comorbid with dyslexia. Comorbidity indicates a medical condition, existing simultaneously, but independently, with another condition \cite{Nicolson2011}.}

\subsection{The Knowledge of Dyslexic Errors}

We decided to use errors written by people with dyslexia as the starting point because they can be used as a source of knowledge. Writing errors of people with dyslexia are not only different from regular spelling errors \cite{Pedler2007}, but are also related to the difficulties that they have \cite{Sterling1998}. Their written errors have been used for various purposes such as studying dyslexia \cite{Lindgren2011}, diagnosing dyslexia \cite{TALE1984}, or for accessibility related purposes such as developing spellcheckers~\cite{Pedler2007}. 

At the same time, one of the main challenges that people with dyslexia face is that they do not consciously detect errors while reading \cite{Bruck1988,WWW2012}. By presenting children with exercises derived by typical dyslexic errors, we aim to stimulate the strategies needed for detecting writing errors and solve them.

\subsection{Content Design}
The game contains 5,000 exercises, 2,500 for English and 2,500 for Spanish. In order to create the exercises, we applied linguistic knowledge and natural language processing techniques. First, we analyzed the errors of two corpora composed of texts written by children with dyslexia in English \cite{Pedler2007} (2,654 errors) and Spanish (1,171 errors) \cite{LREC2014}, finding:

\begin{itemize}
\begin{itemize}
\item[(a)] {\it insertion} of letters, \textit{*arround (around)};
\item[(b)] {\it omission} of letters, \textit {*emty (empty)};
\item[(c)] {\it substitution} of letters, \textit{*scholl (school)}; 
\item[(d)] {\it transposition} of letters, \textit {*littel (little)}; 
\item[(e)] {\it word boundary errors} such as split words, {\it *mis understanding (misunderstanding)}, and run-ons, {\it *alot (a lot)}; and
\item[(f)]  {\it morphology errors} such as using wrong suffixes or prefixes, {\it *warnment (warning)}. 
\end{itemize}
\end{itemize}

As a result of this analysis, we extracted a set of linguistics patterns that occur in the errors. Then, we used these patterns to design the exercises, following the next four steps.

\begin{itemize}
\item[(1)]{\bf Exercises.}
The game presents six types of word exercises according to the six kinds of errors that appear in the analyzed texts: {\it add a letter}; {\it remove a letter}, {\it change a letter}, {\it put the letters in order},  {\it split into words}, and {\it choose the correct word ending}. One example for each case can be found in Figures~\ref{dysapp} and \ref{sopa} (right). For instance, since {\it *alot (a lot)}  is a recurrent word run-on error, there is a split exercise for this case, as shown in Figure \ref{dysapp} (c). 

\item[(2)]{\bf Modification of the Target Words.}
For the exercises that are not derived directly from the incorrect words in the corpora, we apply the linguistic patterns extracted from the errors to the most frequent words. This way, we cover frequent words that might not appear in our corpora.  For instance, when the sound /\textschwa/ (schwa) is represented by the diphthong $<$ou$>$ ({\it tremend\textbf{\textit{ou}}s}) and the triphthong $<$eou$>$ ({\it gorg\textbf{\textit{eou}}s} or {\it courag\textbf{\textit{eou}}s}), these two groups of letters are frequently mistaken between themselves or by other letters such as $<$uo, u, euo$>$. We use these error patterns and apply it to other target words with the same linguistic features, to create new exercises, see Figure \ref{dysapp} (a). 

\item[(3)]{\bf Selection of the Distractors.}
We selected distractors for each exercise word. Distractors are incorrect options in a multiple-choice answer, which resemble the correct option to `distract' the player \cite{mitkov2009semantic}. For instance, similar letters representing similar sounds, such as the occlusive consonants $[$d,b,p,g,t$]$, tend to induce more errors, so we use them as distractors, as in Figure \ref{dysapp} (d).

\item[(4)]{\bf Difficulty Levels.}
The game has five difficulty levels:  {\it Initial},  {\it Easy}, {\it Medium}, {\it Hard}, and {\it Expert}. The levels of the exercises were designed by considering the difficulties of people with dyslexia \cite{Gregor2000}. They have more difficulties with less frequent and longer words, phonetically and orthographically similar words, and words with complex morphology. Hence, in higher difficulty levels, the target word is less frequent, longer, has a more complex morphology, and has a higher phonetic and orthographic similarity with other words.\footnote{We computed the phonetic and orthographic similarity of the words taking into consideration their number of neighbors in each language. That is, words with the same length as the target word which differ in only one letter. That is, Hamming distance one.}
\end{itemize}


\subsection{Implementation}
\sloppy
The application was done in Objective-C by using the Model-View-Controller pattern and a high level abstraction to make it easily portable from iOS\footnote{Search `Dyseggxia' in Apple's App Store and `Piruletras' for its Spanish version.}
to Android\footnote{\small{\url{http://dyseggxia.com/download}}} and later to any other platform as needed. 

\subsection{Interface Design}
Since text presentation has a significant effect on reading performance of dyslexic readers, the interface of {\it DysEggxia} implements the guidelines that -- according to the latest findings in accessibility research \cite{Gregor2000,Kurniawan2006,W4A2012} -- ensures best on-screen text readability for this target group. Text is presented in black on creme background, using \emph{Helvetica}  \cite{ASSETS2013} and a minimum font size of 18 points \cite{W4A2013Wiki}.

To ensure usability and user engagement, we iterated the application in a series of pilot tests with 12 children, using the think-aloud method. For instance, we replaced the written explanations of the exercises by animations and symbols, so no reading is required. Also, to increase long-term engagement, we added in-game achievements by solving different challenges: a penguin appears, grows, and wins prizes. The achievements can be shared via the iOS' Game Center. 


\section{Evaluation}
To study the effect of doing error-based exercises on the spelling skills in Spanish, we conducted an experiment in a primary school. For eight weeks, 48 children had to play {\it DysEggxia} or {\it Word Search}, a word-puzzle game which served as a control condition. Using a within-subject design, we compared the evolution of their reading and writing performance as well as their subjective perceptions. 

We raise the following hypotheses:
\begin{itemize}
	\item \textbf{H.1} Doing exercises based on dyslexic errors helps children with dyslexia to improve their spelling skills.
	\item \textbf{H.2} Doing exercises based on dyslexic errors helps children with dyslexia to improve their reading skills.
	\item \textbf{H.3} Doing exercises based on dyslexic errors rises the subjective perception of reading and spelling skills of children with dyslexia. 
\end{itemize}

\subsection{Participants}
A first group of 54 potential participants with literacy difficulties was selected by their school teachers at School Lestonnac Barcelona. Then, to find out which children had  dyslexia, this group was filtered using a standard test, {\it TALE} \cite{TALE1984}, to diagnose dyslexia in Spanish. {\it TALE} analyzes both reading and writing skills. On the basis of the {\it TALE} scores, we selected 48 participants (29 girls and 19 boys). All of them were diagnosed with dyslexia, with ages ranging from 6 to 11 years ($\bar{x}$= 8.79, $s$ = 1.44).  Using the {\it TALE} scores, we split them into two age groups of the same size according to their literacy skills, which coincidentally matched perfectly with their school year. All but two participants had Spanish as their native language and five of the participants had attention-deficit hyperactivity disorder which is comorbid with dyslexia. All the children had experience playing computer games; 44 of them regularly played games on mobile devices at home.

\subsection{Design}

\begin{figure}[tb!]
	\centering
	\includegraphics[width=8.5cm]{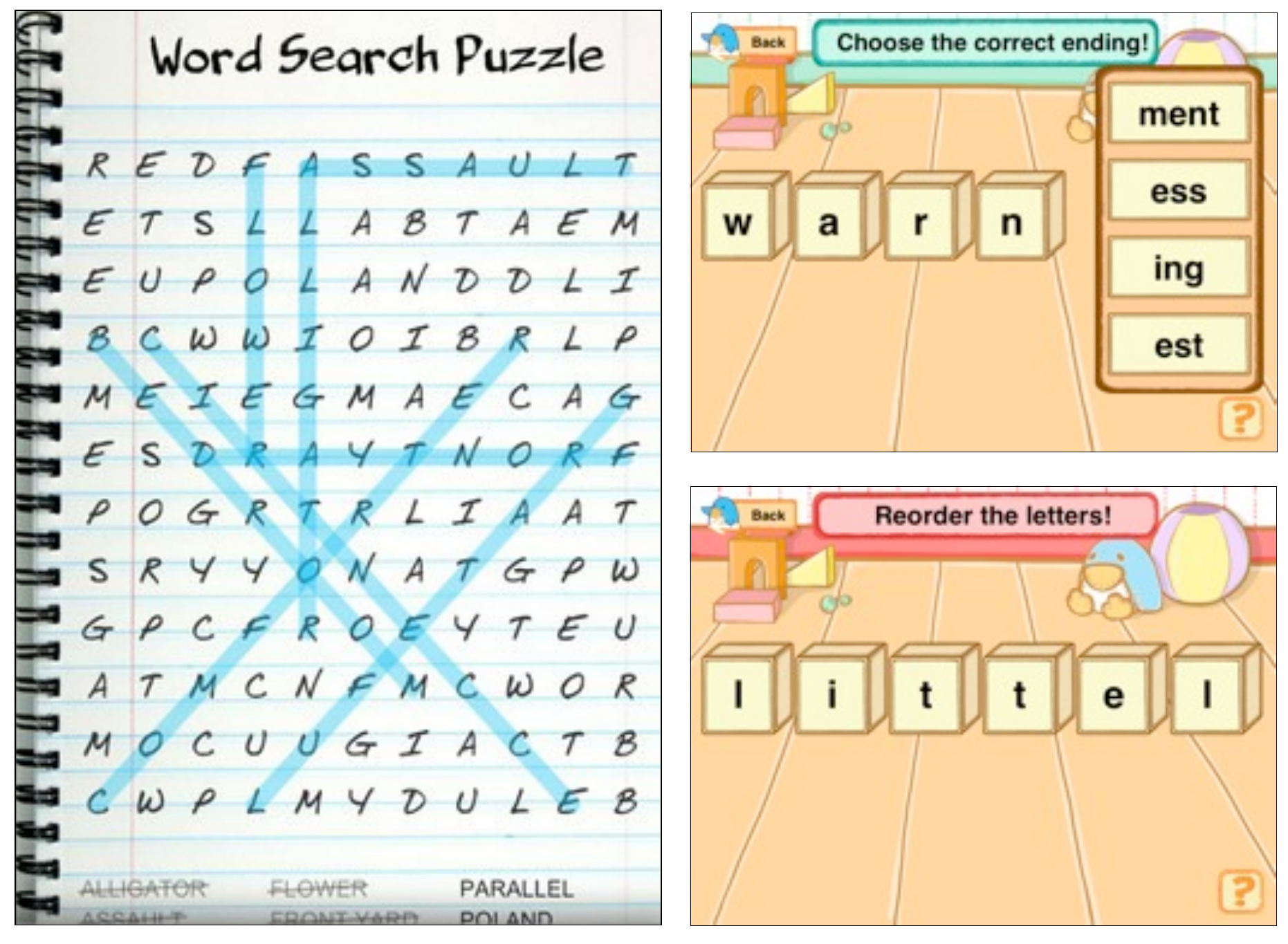}
	\vspace{-0.5cm}
	\caption{{\it Word Search} (left) and {\it DysEggxia} (right).}
	\label{sopa}
	\vspace{-0.3cm}
\end{figure}

The game played served as {\it independent variable} with two conditions. In the experimental condition, the children played {\it DysEggxia} while {\it Word Search}\footnote{Search `Word Search' (`Sopa de Letras' in Spanish) in Apple's App Store.}
served as control condition. Word Search is another game for iPad where the player has to find and mark words hidden in a 
matrix of letters (Figure \ref{sopa}, left).  
We chose {\it Word Search} as a baseline because it was the most similar game to {\it DysEggxia} (see Figure \ref{sopa}, left) that we found in the Apple's App Store.  It even shares many words used in our exercises and both games offered engaging elements, such as the
possibility of accumulating points while playing. Hence, we are able to isolate, whether observed effects are caused due to the specific design of the exercises of {\it DysEggxia}, or simply by playing word-based games on an iPad.

We did not use other specific games for treating dyslexia such as {\it GraphoGame} because their goal is the phonological acquisition in the early stages of the language acquisition while {\it DysEggxia} aims at writing skills of older children. Besides, we could not find such specific games for Spanish. We did not implement a similar game to {\it DysEggxia} containing exercises based on other types of errors because dyslexic spelling errors overlap with regular spelling errors \cite{Sterling1998}.

We used a within-subject design, that is, each participant contributed to both conditions. We divided each age group into two test groups, called A and B. To avoid biases from individual differences, test groups were balanced in terms of reading and writing level, age, and gender. A pre-test was administrated to all the children. Then, group A started by playing {\it DysEggxia} while Group B used {\it Word Search}. After four weeks (12 sessions of 20 minutes each), the first post-test was administrated and the groups switched the games. The length of the second period was the same as the first and thus both groups were exposed to both games for the same amount of time. After the second phase the participants undertook the last post-test. We used this design to cancel out sequence effects.


\subsection{Dependent Measures}
For quantifying writing and reading performance, we used four {\it dependent measures} extracted from the tests. Then, to collect the subjective perceptions of the participants we used subjective ratings extracted from questionnaires.  

As \emph{writing error} we counted the types of writing errors presented in Section 4.2. If a word contained more than one error, they were counted individually. For instance, {\it *litel (little)} has two errors. We used the Levenshtein distance\footnote{The minimum number of single-character edits (insertion, deletion, substitution) required to change the wrong word into the correct word \cite{lev65}.} to compute the number of errors. As Pedler \cite{Pedler2007}, we use Damerau's variant \cite{Damerau1964} where a transposition error counts as a single error. Word boundary and morphology errors counted also as one error even if several letters or words are involved in the error. The number of writing errors depend on the length of the text, this is why we do not directly report the plain number of writing errors. Hence, we compute the following dependent variables:

\textbf{Writing} | {\it Rate of Words with Errors}. The number of words with at least one writing error divided by the total number of words.

\textbf{Writing} | {\it Rate of Errors per Word}. An incorrect word can have more than one writing error, {\it*probley (probably)}. Therefore, we also compute the average rate of errors per word. We define it as the number of writing errors divided by the number of total words.

\textbf{Writing} | {\it Rate of Errors per Wrong Word}. This measure reports the sum of all errors divided by the number of words with at least one error. That is, it measures the severity of the incorrectness of a wrong word.

\textbf{Reading} | {\it Rate of Errors per Word}. We define it as the number of aloud reading errors divided by the number of total words. We take into account the same types of errors as before. In the same way, omitted and added words counted as one error.

\textbf{Subjective Reading and Writing Skills}. Subjective reading and writing skills are measured by self-report questionnaires. On a 5-point Likert scale, the participants had to rate how well they think they can read and write. The scores range from 1 = \emph{Very bad} to 5 = \emph{Very good}. 


\subsection{Materials}
\vspace{0.1cm}


All the participants took three tests --one pre-test and two post-tests-- to measure the development of reading and writing skills during the experiment. 

\begin{itemize}
\item[(a)] {\bf Test Structure.} Each test was composed of two parts: a dictation (writing task) and a part to read out loud (reading task). Each of the tasks contained two sentences and a list of eight words.
While two sentences and eight words may appear little, the pedagogues insisted limiting the tests to this size to make them comparable to the ones that they do in class.
Even if the games use only single words, we include both sentences and single words in the tests because single words are processed differently when they are used individually than when they are inserted into sentences \cite{Pinker2009}.
By using both the list and the text, we can test whether the exercises based on single words can also impact at the sentence level.


\item[(b)] {\bf Word Content.} For the experiment, we created three texts for the writing test and three texts for the reading test. To ensure that the participants did not just remember words from previous tests, the lexical words used in the texts were different, with the exception of functional words (prepositions, conjunctions, pronouns, and articles), such as {\it at}, {\it in}, or {\it the}.
Each test contained twelve words that appeared in the {\it DysEggxia} exercises. For the rest of the test, we used words that appear in the children's school textbooks.

\end{itemize}

\subsection{Procedure}
\begin{figure}[tb!]
	\centering
	\includegraphics[width=8.5cm]{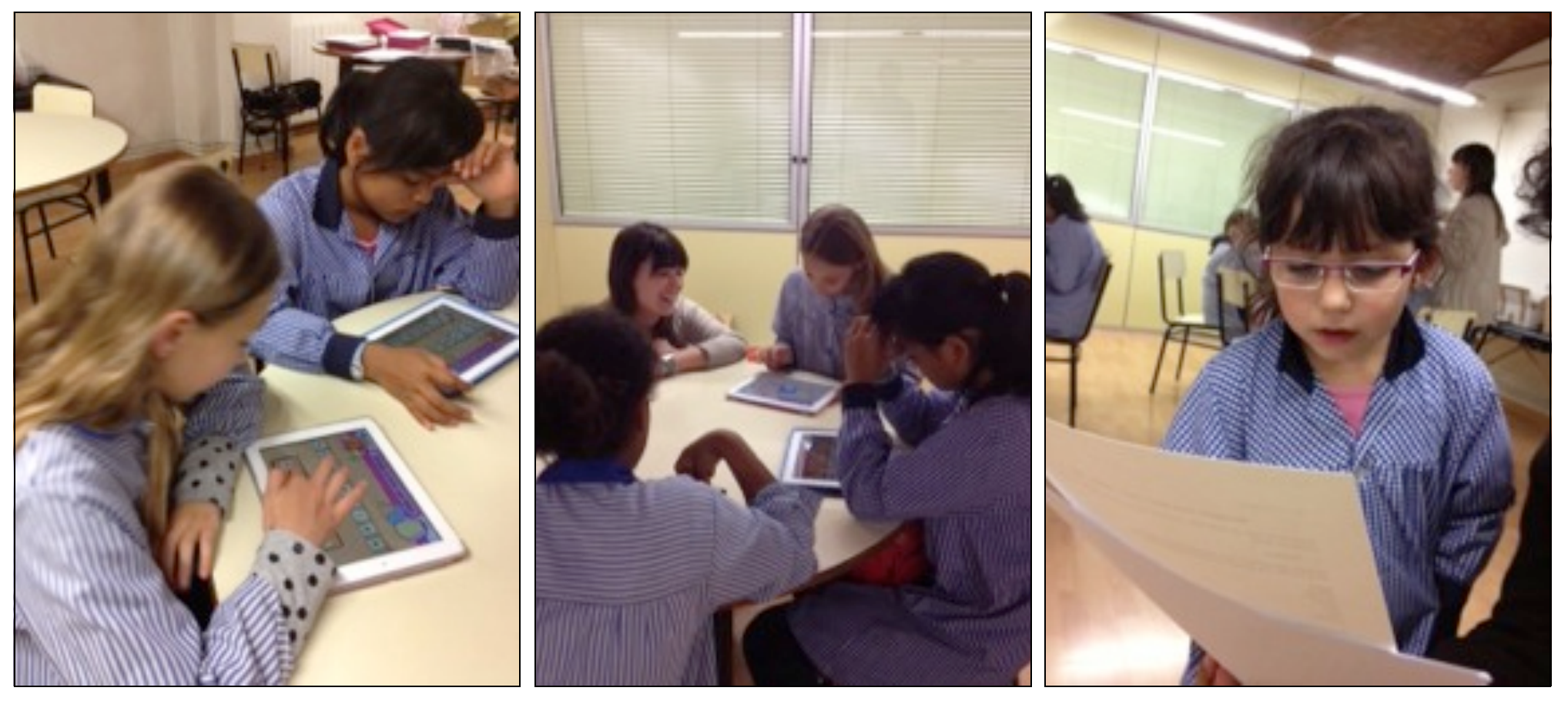}
		\vspace{-0.5cm}
	\caption{Experimental sessions (left and centre) and reading test (right). Photos included with the parents' permission.}
	\label{fig:procedure}
		\vspace{-0.3cm}
\end{figure}

To motivate the children to give their best, we introduced the study as a contest.
We announced prizes, to be awarded in a ceremony at the end of the study, for the players who reached the highest scores summing the points of both games. 

First, we administered a questionnaire to collect demographic information: age, gender, school year, their native languages, and their habits of playing games in mobile devices. Then, we conducted the reading and writing pre-tests prior to exposing the children to any of the conditions and issued the subjective-skills questionnaires. 

Recall that in Phase 1, Group A played {\it DysEggxia} and Group B played {\it Word Search}. We sent a letter to the parents, asking them to make sure that the children did not play the games at home. After 12 sessions/4 weeks, Phase 1 ended with the first post-test (Figure \ref{fig:procedure}, right). This test comprised the reading and writing tests and the subjective-skills questionnaire and allowed us to compute how much the children had improved in either of the conditions.

\begin{table*}[t!]
\begin{center}
\begin{tabular}{llll|llll} 
	\toprule
	{\bf Dependent Variable} &  \multicolumn{3}{l}{{\it DysEggxia}} & \multicolumn{3}{l}{\it Word Search} & Significance \\
	\midrule
	{}   & Pre   & Post 	& Change (SD)  & Pre   & Post 	& Change (SD) & {}\\
	\\
	Writing / Words With Errors	&  28.9\% & 23.9\% & -5.0 (17.3) pp  &  28.2\% & 25.9\% & -2.3 (15.6) pp & $p = 0.355$\\
	Writing / Errors per Word   &  0.360 & 0.288 & -0.072 (0.242)  &  0.325 & 0.355 & +0.030 (0.223) & $p = 0.029$\\
	Writing / Errors per Incorrect Word   &  1.178 & 1.007 & -0.171 (0.559)  &  1.080 & 1.244 & +0.164 (0.487) & $p = 0.011$\\
	\\
	Reading / Errors per Word    &  0.117 & 0.106 & -0.011 (0.093)  &  0.129 & 0.114 & -0.015 (0.136) & $p = 0.410$\\
	\\
	Subjective Writing Skills   &  3.395 & 3.721 & +0.326 (1.190)  &  3.326 & 3.581 & +0.256 (0.978) & $p = 0.426$\\
	Subjective Reading Skills  &  3.488 & 3.933 & +0.445 (1.099)  &  3.535 & 3.767 & +0.233 (0.922) & $p = 0.176$\\	
	\bottomrule
\end{tabular}
\end{center}
\vspace{-0.2cm}
\caption{The numbers show mean across all children by condition (pp = percentage points).}
\label{tab:resultsall}
\vspace{-0.1cm}
\end{table*}%

In Phase 2, the conditions were reversed: for another 12 sessions / 4 weeks, Group A played {\it Word Search} and Group B played {\it DysEggxia}. The phase was concluded by the second post-test, containing the reading and writing tests and a subjective-skills questionnaire, which allowed us to compute the level of improvement for each child in this phase.

The sessions were conducted at School Lestonnac Barcelona. During lunch breaks the children went into a quiet classroom, where they played the game that they were assigned to for 20 minutes (Figure \ref{fig:procedure}, left and centre). While doing the exercises, one psychologist and one pedagogue constantly supervised the children. If a child got stuck in an exercise, they helped to avoid frustration (Figure \ref{fig:procedure}, centre). Finally, all children were gathered in a big classroom to celebrate the end of the contest. The winners of each age group were awarded with a diploma. All children received a token gift for their participation, namely crayons and a small toy.

\section{Results}
In the first step we cleaned up the data. We had to omit the data of five children. One child refused to take part in the tests. One child had played {\it DysEggxia} while she should have played {\it Word Search}, which violates the assumptions of the experimental design. Another child had not been schooled until the age of 5 and her general skills turned out not to be comparable with the rest of her group. Two children could not attend all of the tests due to illness, and hence did not contribute to all conditions equally. Thus, our quantitative results reflect the data of the remaining 43 children.

In order to isolate the effect of the two conditions, {\it DysEggxia} and {\it Word Search}, we computed for each child how the values in all dependent variables evolved during the two phases, \emph{i.e.} between tests 1 and 2, and tests 2 and 3. 
Combined, this provides one value per participant per condition, that is 86 data points for each dependent variable. To test for significant effects, we used paired t-tests for parametric data and Wilcoxon-Signed-Rank tests for non-parametric data. We used Shapiro-Wilk tests to determine whether each variable was normally distributed, hence parametric data, or not. Table \ref{tab:resultsall} summarises scores for all dependent variables.


Practicing with {\it DysEggxia} significantly decreased the number of errors that the children made in the writing tests.
There was a significant effect on the number of \emph{errors per word} ($T = 629.5,~p = 0.029,~r = 0.292$) 
and on \emph{errors per incorrect word} ($T = 607,~p = 0.011,~r = 0.353$). There was no significant effect on \emph{words with errors} ($t(42) = 0.935,~p = 0.355$). This means that due to playing {\it DysEggxia}, there were less overall errors. The results are not conclusive whether this causes children also to write more words without any error.

There was no significant effect on the number of reading errors ($T = 427.0,~p = 0.41$). 
This means that {\it DysEggxia} did not have a significant effect on improving reading errors in this experiment.

In addition, we did not find a significant effect on the subjective writing skills ($T = 226.0,~p = 0.426$), nor on the subjective reading skills ($T = 211.5,~p = 0.176$). 
Hence, we are not able to tell whether exercising with {\it DysEggxia} affected the subjective reading and writing skills.

\section{Discussion}
When playing {\it DysEggxia} children improved their spelling significantly compared to playing {\it Word Search}. Hence, we accept {\bf H.1} {\it Doing exercises based on dyslexic errors helps children with dyslexia to improve their spelling skills.} We found a significant reduction of errors after playing {\it DysEggxia} compared to the control condition. The  \emph{rate of errors per word} decreased by 20\%, that is, significantly less overall errors ($p=0.029$). In particular, wrong words contained less errors, e.g. \emph{*acer} instead of \emph{*azer} ({\it `to do'}), the correct word being \emph{hacer} ($p=0.011$), with a 14.5\% decrease. The results are inconclusive whether this will lead to less wrong words, in spite of having a decrease of 17.3\% for this variable. 

We reject hypothesis {\bf H.2} {\it Doing exercises based on dyslexic errors helps children with dyslexia to improve their reading skills}, as we found no significant effects on the improvement of the rate of reading errors, although there was a 9.4\% decrease. There are three possible reasons. First, the exercises were specifically designed to train spelling skills, since they were based on writing errors. Also, we used animations and symbols instead of text to present the instructions of the game. Hence, {\it DysEggxia} requires very little reading beyond the exercise words. 
Second, our game did not include reinforcement exercises that targeted the development of reading skills. For instance, we did not include exercises to reinforce the grapheme-phoneme connections, such as {\it Literate} \cite{Lyytinen2007} does, which leads to an improvement in reading words. 
Finally, {\it DysEggxia} does not have the characteristics of computer action games which were shown to improve the children's attention and hence their skills in decoding words~\cite{Franceschini2013}. 




We reject hypothesis {\bf H.3} {\it Doing exercises based on dyslexic errors rises the subjective perception of reading and spelling skills of children with dyslexia}, as no significant change in the subjective ratings was observed.
This can be explained with the reduced ability of people with dyslexia to identify whether a word is correct or not \cite{Bruck1988,WWW2012}.
There are a number of studies which confirm that the reading performance of people with dyslexia does not correlate with their subjective perception of this performance. For instance, in a study using eye-tracking, the textual layout which lead to the fastest reading speed among people with dyslexia was chosen as the subjective best layout by only half of the participants \cite{W4A2012}. This is one of the reasons of why dyslexia is called a {\it hidden disability} \cite{Lyon2003}, because, as mentioned above, people with dyslexia cannot perceive whether they are reading and/or writing correctly. Also, we did not disclose the results of the tests during the experiment, so the children had no indications on whether they were improving or not.

{\bf Limitations.} Our study has two main limitations. First, our current results are valid only for Spanish. Nonetheless, {\it DysEggxia} also has  exercises in English, which are designed with the same criteria as the Spanish exercises. However, English and Spanish have different orthographies, and manifestations of dyslexia depend on the language orthography, in particular with regard to their grade of consistency and regularity \cite{Brunswick2010}.
English has an opaque --or deep-- orthography, \emph{i.e.} the relationships between letters and sounds are inconsistent (compare {\it vase} with {\it base}). 
Opaque languages present a greater challenge to beginners than languages with transparent --or shallow-- orthography, \emph{i.e.} consistent mapping between letters and sounds, such as Spanish.
Therefore, an additional study would be needed to extend the results to English. 

Second, we did not introduce a third condition, in which the children did not play any word game. Hence, we cannot isolate, to what extent simply playing an iPad word related game already has positive effects with respect to how much a child learns in the same period of time without playing
The reason for not including this condition is because we did not want to frustrate children by leaving them out of the competition for weeks, and make them lose their engagement.
Nevertheless, by selecting another word game as baseline, the identified effects can be clearly attributed to {\it DysEggxia}'s specific design, which is the cause-and-effect relationship we primarily wished to establish.

\section{Conclusions}

We presented a method to train and improve the spelling skills of children with dyslexia. In contrast to previous work, the exercises are based on common errors by children with dyslexia and are presented on a tablet game. Our results show that our method integrated in the game {\it DysEggxia} has an impact in improving writing skills better than doing word games with correct words.

Our game has shown effectiveness and benefits for Spanish early-writers. It provides exercises in English and Spanish, which are, after Mandarin Chinese, the second and third most spoken languages in the world. Also, it is a simple, cost-efficient way of helping children with dyslexia to improve their spelling skills. Indeed, the iOS version app has been installed more than 17,000 times since its release in June 2012. So far, three schools that support children with dyslexia\footnote{ {\it Centro Creix Barcelona}: {\small{\url{www.creix.com/Barcelona}}}\\ {\it Centro Coddia:} {\small{\url{www.coddia.com}}} and \\ {\it Uditta}: {\small{\url{http://www.uditta.com/}}}} have adopted {\it DysEggxia} into their curriculum. Currently, we are working on the adaptation of the game to Catalan and German languages.

Future work also needs to study additional strategies, such as the reinforcement of grapheme-to-phonemes transformations, to not only train writing but also reading skills. We plan to research the effect of different strategies to tailor the exercises on the basis of the child's performance.

\section{Acknowledgements}
We thank the psychologist Cat\'in Sintes and the pedagogue J\'essica Leal for supervising all the sessions in the school.
We thank Azuki G\`orriz for the beautiful design of the graphics used in {\it DysEggxia}.  We acknowledge Col$\cdot$legi  Lestonnac Barcelona for their collaboration in the study. We also thank Nancy Crushen White for her feedback during the first phase of development of the app. Finally, we specially thank Ricardo Baeza-Yates for his invaluable support during all the development process of  {\it DysEggxia} and this article.

\bibliographystyle{abbrv}

\balancecolumns

%
%

\end{document}